# Unveiling Nano-scale Crystal Deformation using Coherent X-ray Dynamical Diffraction

Longlong Wu[1,2*], David Yang[2], Wei Wang[2,3], Shinjae Yoo[4], Ross J. Harder[5], Wonsuk Cha[5], Aiguo Li[1] and Ian K. Robinson[2,6,‡]

*[1]Shanghai Advanced Research Institute, Chinese Academy of Sciences, Shanghai 201210, China*

*[2]Condensed Matter Physics and Materials Science Division, Brookhaven National Laboratory, Upton, New York 11973, USA*

*[3]Department of Physics, University of Science and Technology of China, 230026 Hefei, Anhui, China*

*[4]Computational Science Initiative, Brookhaven National Laboratory, Upton, NY 11973, USA*

*[5]Advanced Photon Source, Argonne National Laboratory, Lemont, Illinois 60439, USA*

*[6]London Centre for Nanotechnology, University College London, London WC1H 0AH, UK*

*Correspondence emails: [*]wull@sari.ac.cn, [†]irobinson@bnl.gov*

## Abstract

**Visualization of internal deformation fields in crystalline materials helps bridge the gap between theoretical models and practical applications. Applying Bragg coherent diffraction imaging under X-ray dynamical diffraction conditions provides a promising approach to the longstanding challenge of investigating the deformation fields in micron-sized crystals. Here, we present an automatic differentiation-based reconstruction method that integrates dynamical scattering theory to accurately reconstruct deformation fields in large crystals. Using this forward model, our simulated and experimental results demonstrate that three-dimensional local strain information inside a large crystal can be accurately reconstructed under coherent X-ray dynamical diffraction conditions with Bragg coherent X-ray diffraction imaging. These findings open an avenue for extending the investigation of local deformation fields to microscale crystals while maintaining nanoscale resolution, leveraging the enhanced coherence and brightness of advanced X-ray sources.**

## Introduction

Visualization of internal deformation fields in functional crystal materials is fundamentally important for their design and practical applications[1,2]. Such visualization facilitates our understanding of strain-related microscopic mechanisms in these small crystals, as their specialized functionality increasingly relies on complex local deformation information that deviates from an ideal crystal structure. This is especially true for microscale crystal materials, such as ferroelectrics[3], piezoelectrics[4], shape memory alloys[5], and semiconductor microcrystals[6], which exhibit unique behaviors due to their intermediate scale that bridges the gap between nanoscale and bulk materials. Advances in deformation detection technology are driving the need for the development, optimization, and application of these functional crystal materials, for example in mobility engineering of semiconductors[7] or control of electronic band structure in quantum materials[8].

Among the few powerful X-ray crystal microscope methods capable of measuring internal three-dimensional deformation fields, Bragg Coherent Diffraction Imaging (BCDI) offers great potential for investigating various length scales of crystals with nanoscale resolution[9], based on coherently diffracted X-ray patterns. In a typical BCDI experiment, an intense coherent X-ray beam is used to record interference patterns in the vicinity of one or several Bragg reflections of an illuminated crystal[10-12]. However, the phase information of these patterns is lost during measurement. The inversion of these patterns using traditional iterative methods[13,14] provides a high-resolution projection of the local deformation field along the reciprocal lattice vector of the crystal, which is obtained from the phase of the recovered complex function[12]. This process relies on three-dimensional Fourier transformations between real and reciprocal space, which confines the analysis to the framework of kinematical approximation.

As a simplification of the more rigorous dynamical diffraction theory[15], only the interactions between the primary incident X-ray beam and the crystal atoms are considered in the kinematical approximation, while higher-order effects are neglected. Consequently, in most BCDI experiments, the crystal size has been limited to nanoscale dimensions to satisfy the kinematical approximation and minimize artifacts in the reconstructed deformation fields caused by dynamical X-ray diffraction effects. As fourth-generation synchrotron radiation sources and X-ray free-electron lasers are developed worldwide[16,17], the significant increase in coherence length of X-rays enables the experimental probing of deformation in larger crystals. While BCDI offers a unique capability to resolve three-dimensional (3D) strain in thick crystals with nanoscale resolution, the absence of a robust method to account for dynamical diffraction effects has hindered the quantitative study of deformation fields in crystals whose size exceeds the extinction depth or approaches the *Pendellösung* distance[18,19]. To expand these opportunities in recent years, there has been growing interest in understanding the role of dynamical diffraction effects in BCDI experiments, and several approaches have been proposed to address refraction and absorption effects arising from coherent dynamical diffraction[15,20-22] while a strong coupling between the incident wave and diffractive wave has generally been ignored. Consequently, these methods are only applicable under limited conditions[23].

In this paper, we demonstrate an automatic differentiation-based approach that reconstructs the internal deformation field of a crystal from its experimental Bragg coherent diffraction patterns using a coherent X-ray dynamical diffraction formalism. This model is implemented through the integration of the Takagi-Taupin equations (TTEs) on an orthogonal grid during the BCDI reconstruction process. In our approach, the TTEs are solved using exponential Rosenbrock-type methods for numerical integration. By incorporating this dynamical diffraction model into a mini-batch stochastic gradient descent optimization framework, the deformation field of the finite crystal is then reconstructed from its corresponding 3D Bragg coherent diffraction patterns. We have validated our approach using both simulated and experimental data. Compared with traditional iterative methods, our approach enables the direct incorporation of dynamical diffraction theory without compromising the accuracy of the determined internal lattice distortion field. It also provides a more reliable and precise path to convergence. By employing BCDI reconstruction based on dynamical diffraction theory, our approach not only provides a way to investigate deformation fields in microscale crystals but also enables the utilization of the enhanced coherence properties offered by advanced X-ray sources.

## Results

### Dynamical diffraction and computational model

Figure 1a illustrates the experimental setup for a typical BCDI measurement, based on the geometry at beamline 34-ID-C at the Advanced Photon Source, Argonne National Laboratory. It specifically shows the diffraction geometry within the standard laboratory coordinate system $\mathrm{X'Y'Z'}$. A small crystal is fully illuminated by a coherent X-ray beam with wavevector $\mathbf{k}_0$, with the crystal size being smaller than the beam's coherence length. The incident X-ray wavefront is assumed to be a plane wave. The diffracted wavevector at a position on a far-field two-dimensional (2D) detector is defined as $\mathbf{k}_\mathrm{f}$, which equals to $\mathbf{k}_h$ when satisfying the Bragg condition for a reciprocal lattice vector $\mathbf{h}$, *i.e.*, $\mathbf{h} = \mathbf{k}_h - \mathbf{k}_0$. Both $\mathbf{k}_0$ and $\mathbf{k}_\mathrm{f}$ vectors are defined in vacuum and have same magnitude. Here, it should be noticed that $\mathbf{k}_\mathrm{f}$ is a variable vector, and $\mathbf{k}_\mathbf{h}$ is a constant vector, determined solely by $\mathbf{h}$ and $\mathbf{k}_0$. The momentum transfer vector $\mathbf{Q}$ is therefore given as $\mathbf{Q} = \mathbf{k}_\mathrm{f} - \mathbf{k}_0$, which is a variable vector and not the same as the fixed point $\mathbf{h}$. To measure the Bragg coherent diffraction pattern around a reciprocal lattice vector $\mathbf{h}$, which requires the satisfaction of the Bragg condition, *i.e.,* when $\mathbf{Q} = \mathbf{h},$ the 2D detector is positioned using a combination of the horizontal $\delta$ and vertical $\gamma$ angles and the crystal sample is rotated around the Y′-axis by angle $\theta$. Since the diffracted coherent diffraction pattern recorded by the detector is a 2D slice in the reciprocal space near $\mathbf{h}$, the 3D diffraction volume in reciprocal space is sampled by a stack of 2D slices by rotating the crystal sample in small $\Delta\theta$ steps along the rocking curve direction. In this way, the 3D diffraction pattern is measured slice by slice within a parallelepiped volume of Fourier space. When rotating the crystal by the angle $\Delta\theta$, the reciprocal lattice vector is changed by $\Delta\mathbf{q} = \mathbf{h}' - \mathbf{h}$, where $\mathbf{h}'$ is the updated reciprocal lattice vector at rotated crystal orientation. Figure 1b presents a detailed 2D sketch of the sampling system, introducing the XYZ coordinate system. It is viewed from the scattering plane defined by $\mathbf{k}_0$ and $\mathbf{k}_h$ vector.

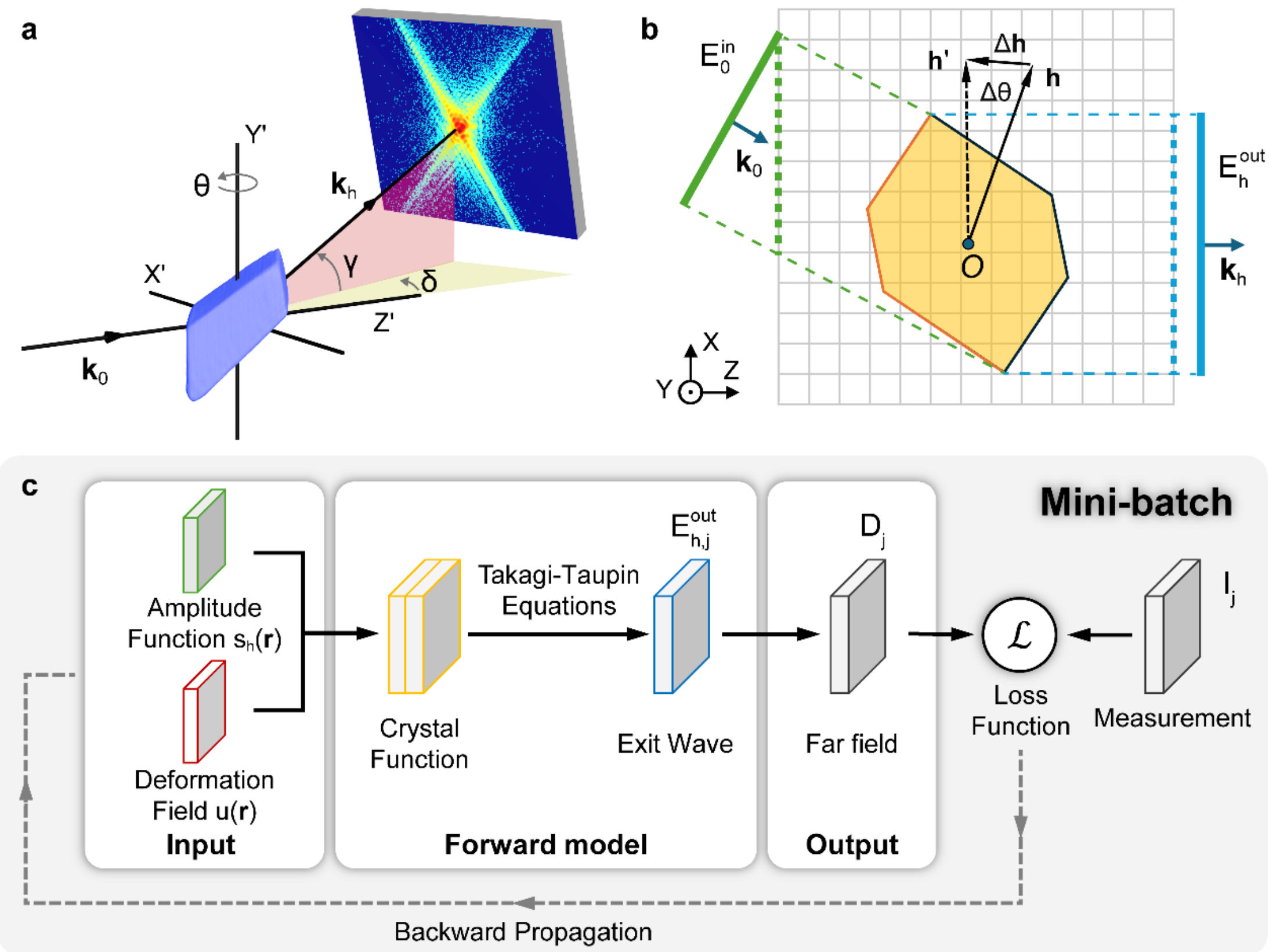

**Fig. 1 | Computational graph for coherent X-ray dynamical diffraction.** (a) Basic geometry of a BCDI experiment. (b) Schematic of coherent X-ray diffraction from a crystal in BCDI experiments. (c) Computational graph for the reconstruction. The forward path marked by solid arrows predicts the diffraction pattern measured by the X-ray

detector and evaluates the loss function. The backward path marked by dashed arrows computes the gradients with respect to each optimizable variable.

The mathematical foundation of BCDI within the framework of the kinematical diffraction approximation can be formulated as a three-dimensional Fourier transform based on the 3D diffraction pattern. This approximation becomes invalid when the measured crystal size is close to its *Pendellösung* distance in Laue geometry (or extinction distance in Bragg geometry). In such cases, the more rigorous dynamical diffraction theory should be employed. According to dynamical X-ray diffraction theory, under the two-beam diffraction approximation, the wavefield inside a crystal can be expressed as a sum of modulated plane waves:

$$\mathbf{E}(\mathbf{r}) = E_0(\mathbf{r})e^{i\mathbf{k}_0\cdot\mathbf{r}} + E_h(\mathbf{r})e^{i\mathbf{k}_h\cdot\mathbf{r}}, \quad (1)$$

where $\mathbf{k}_0$ and $\mathbf{k}_h$ represent the wavevectors of the transmitted and diffracted waves, respectively. The complex wavefronts for the transmitted $E_0(\mathbf{r})$ and diffracted $E_h(\mathbf{r})$ waves are the solutions of the well-known Takagi-Taupin Equations (TTEs)[18,19,24,25]. For Bragg coherent X-ray diffraction imaging, they can be obtained by solving the symmetric version of two beam TTEs[15], which are given by:

$$\begin{aligned} 2i(\mathbf{k}_0\cdot\nabla)E_0(\mathbf{r}) &= k^2\left[\chi_0 E_0(\mathbf{r}) + C\chi_{\bar{h}}E_h(\mathbf{r})\mathrm{e}^{-i\Delta\mathbf{q}\cdot\mathbf{r}+i\mathbf{h}\cdot\mathbf{u}(\mathbf{r})}\right], \\ 2i(\mathbf{k}_h\cdot\nabla)E_h(\mathbf{r}) &= k^2\left[\chi_0 E_h(\mathbf{r}) + C\chi_h E_0(\mathbf{r})\mathrm{e}^{i\Delta\mathbf{q}\cdot\mathbf{r}-i\mathbf{h}\cdot\mathbf{u}(\mathbf{r})}\right]. \end{aligned} \quad (2)$$

Here, $\chi_0$ represents the average electric susceptibility of the crystal, with its real and imaginary parts describing refraction and absorption effects, respectively. $\chi_{h,\bar{h}}$ are Fourier coefficients of the susceptibility function of the crystal for the diffracted beams, which are related to the crystal structure factor at the corresponding reciprocal lattice vectors, as denoted by their subscripts. The symbol $C$ stands for the polarization factor, which equals to $\cos 2\theta_\mathrm{B}$ in the case of a π polarization and unity in the case of a σ polarization. Here, $\theta_\mathrm{B}$ is the corresponding Bragg angle. Since we assume that the wavevectors $\mathbf{k}_0$ and $\mathbf{k}_h$ are the same inside and outside of the crystal[15,18,19,24], the $E_0(\mathbf{r})$ and $E_h(\mathbf{r})$ are continuous on the boundary of the crystal. It makes the Eq. (2) particularly convenient in the case of a 3D crystal with arbitrary shape. Consequently, Fourier Components $\chi_0$, $\chi_h$ and $\chi_{\bar{h}}$ of susceptibility will decrease to zero outside the crystal. To account for this, they are multiplied by the introduced crystal amplitude function $s_{0,h,\bar{h}}$, *i.e.*, $\chi_{0,h,\bar{h}}$=$s_{0,h,\bar{h}}(\mathbf{r})\chi_{0,h,\bar{h}}$. In the case of a perfect crystal, $s_{0,h,\bar{h}}$ follow the same distribution, where they equal unity within the crystal and zero outside the crystal. However, it should be noticed that in the presence of the deformation field, $s_{h,\bar{h}}$ are related to modulations of atomic planes associated with the measured Bragg reflection of the crystal and $s_0$ is proportional to the electron density of the crystal. $\mathbf{u}(\mathbf{r})$ is the local deformation field that describes the displacement of atoms from their ideal positions. If the coupling between transmitted and diffracted waves and the effects of refraction and absorption in the Eq. (2) (see SI for details) are neglectable, Eq. (2) can be solved analytically. Then, the far-field Bragg coherent diffraction intensity can be estimated by Fourier transformation of the crystal function $S_h(\mathbf{r})$, defined as $S_h(\mathbf{r}) = s_h(\mathbf{r})\mathrm{e}^{i\varphi_h(\mathbf{r})}$. Here, the phase by its definition is given as $\varphi_h(\mathbf{r}) = -\mathbf{h}\cdot\mathbf{u}(\mathbf{r})$. This estimation corresponds to the kinematical diffraction condition.

Generally, Eq. 2 can only be solved numerically[26,27]. One traditional approach adopts an oblique coordinate system to solve the TTEs, with the axes aligned with the incident and diffracted wavevectors. To converge, it requires an iterative solution procedure. When the scattering simulation interfaces with material models that require orthogonal grids, the use of oblique coordinates necessitates extra interpolation steps. However, solving the TTEs directly in a orthogonal coordinate system with a finite difference scheme is equivalent to linear interpolation and will lead to accumulated interpolation errors at each step[26]. Thus, by solving the TTEs in reciprocal space and implicitly utilizing Fourier interpolation, the approach we present here uses an orthogonal grid but avoids these errors (see SI for details). When solving the TTEs in an orthogonal grid, the orientation of the grid can be arbitrary. For a BCDI experiment, as shown in Fig. 1(b), it can be generally assumed, that $\mathbf{k}_0 \cdot \hat{z} > 0$ and $\mathbf{k}_\mathrm{h} \cdot \hat{z} > 0$, where $\hat{z}$ is the orthonormal unit vector is along the Z axis. To make the calculation of the diffracted wavefront more convenient, as shown in Fig. 1(b), the origin of the coordinate system is chosen based on the scattering plane, aligned within the XOZ-plane, where $\mathbf{k}_\mathrm{h}$ is along the z-axis. Therefore, by applying the transverse Fourier transform (*i.e.,* the 2D Fourier transformation operation $\mathcal{F}_\perp$ within the XOY plane) to Eq. (2), the TTEs can be rewritten as:

$$\begin{aligned}\frac{\partial}{\partial z}\tilde{E}_0(\mathbf{q}_\perp, z) &= -\frac{i2\pi}{k_{0,\mathrm{z}}}\mathbf{q}_\perp \cdot \mathbf{k}_{0,\perp} \times \tilde{E}_0(\mathbf{q}_\perp, z) - \frac{ik^2}{2k_{0,\mathrm{z}}}\mathcal{F}_\perp\{C\chi_{\bar{h}}E_h(\mathbf{r})\mathrm{e}^{-i\Delta\mathbf{q}\cdot\mathbf{r}+i\mathbf{h}\cdot\mathbf{u}(\mathbf{r})} + \chi_0 E_0(\mathbf{r})\},\\ \frac{\partial}{\partial z}\tilde{E}_h(\mathbf{q}_\perp, z) &= -\frac{i2\pi}{k_{h,\mathrm{z}}}\mathbf{q}_\perp \cdot \mathbf{k}_{\mathrm{h},\perp} \times \tilde{E}_h(\mathbf{q}_\perp, z) - \frac{ik^2}{2k_{h,\mathrm{z}}}\mathcal{F}_\perp\{C\chi_h E_0(\mathbf{r})\mathrm{e}^{i\Delta\mathbf{q}\cdot\mathbf{r}-i\mathbf{h}\cdot\mathbf{u}(\mathbf{r})} + \chi_0 E_h(\mathbf{r})\}.\end{aligned} \tag{3}$$

Here, the subscripts $\perp$ are used to represent the XOY-plane component of vectors and the subscript z represents the corresponding perpendicular z component. Tildes represent the transforms of functions, *i.e.,* $\tilde{E}_0 = \mathcal{F}_\perp(E_0)$ and $\tilde{E}_h = \mathcal{F}_\perp(E_\mathrm{h})$. $\mathbf{q}$ is the deviation vector from the exact Bragg condition, where $\mathbf{q} = \mathbf{Q} - \mathbf{h}$. Variations of the amplitude function inside the crystal describe modulations of atomic planes associated with the chosen reflection **h** and not electron density modulations. Then, the complex wavefront for the transmitted $E_0(\mathbf{r})$ and diffracted $E_\mathrm{h}(\mathbf{r})$ can be obtained by solving the above equations with Exponential Runge-Kutta methods[26]. At the jth rotation angle, the coherent X-ray diffraction patterns $D_j(\mathbf{q})$ can be then approximated by propagating the exit wavefront of diffracted $E_{h,j}^{\mathrm{out}}(\mathbf{r})$ to the far-field, which can be evaluated using the squared Fourier transform magnitude $D_j(\mathbf{q}) = \left|\mathcal{F}_\perp\left[\mathrm{E}_{h,j}^{\mathrm{out}}(\mathbf{r})\right]\right|^2$. While Eq. (3) is only valid for a pure σ or π polarization, it can be still applied for a general diffraction geometry. To control the effect of polarization, one can either tune the polarization direction of the incident x-ray to make the geometry into a pure σ or π polarization or, alternatively, use an appropriate sample geometry. As the TTEs are evaluated on an orthogonal grid, Fourier interpolation is implicitly utilized at the level of the individual finite difference step. We use the ability to freely choose the computational grid to make the implementation of this approach easier. We implemented this approach within automatic differentiation framework[28] for dynamical BCDI reconstruction. Figure 1(c) shows the corresponding schematic of the computational graph for the proposed approach to BCDI reconstruction, where the amplitude function $s_h(\mathbf{r})$ and deformation field $u(\mathbf{r})$ of a measured crystal are directly optimized. As a nonlinear optimization problem, the model evaluates the difference between the measured $I_j$ and the corresponding calculated $D_j$ to find the minimum of a loss function. Then, the partial derivatives of the loss function with respect to each parameter are backpropagated to update these parameters. For computational efficiency, we adopted a mini-batch gradient descent strategy to find the minimum of the loss function[29], where a 2D subset of the 3D input coherent X-ray diffraction dataset (*i.e.,* less than the full dataset) was used at each update until all the measured 2D coherent diffraction patterns had been processed. The 3D dataset was divided into five mini-batches with random order after each epoch. During each mini batch, the target variables were updated with each input subset. The best match between theoretical and experimental diffraction patterns can only be achieved if the amplitude function and deformation field are determined as accurately as possible.

For a coherent X-ray diffraction measurement, the finite X-ray photon statistics and inevitable noise in the measured coherent X-ray diffraction patterns can lead to artifacts in the reconstructed object[29]. Therefore, an adaptive total variation[30-32] (ATV, *i.e.,* $L_p$ norm to the power of $q$) denoising was also applied in our proposed BCDI reconstruction method under dynamical X-ray diffraction conditions. The final loss function is a combination of negative maximum likelihood estimation[29] and ATV, averaged over all the rocking positions in a mini-batch $l$ with size of $L$, formulated as:

$$\mathcal{L} = \frac{1}{L} \sum_{j \in l} \ell_j\left(D_j, I_j\right) + \alpha \cdot ATV(S_h), \tag{4}$$

Here, the maximum likelihood estimation is given as $\ell(\mathbf{q}) = \frac{1}{M}\Sigma\left[D_j^{\frac{1}{2}}(\mathbf{q}) - I_j^{\frac{1}{2}}(\mathbf{q})\right]^2$ and $M$ is the total number of pixels in a diffraction pattern. $\alpha$ is a weight parameter. $\mathrm{S_h}$ is the crystal function given as $S_h(\mathbf{r}) = s_h(\mathbf{r})\mathrm{e}^{i\varphi_h(\mathbf{r})}$. ATV is given as $ATV(S_h) = \frac{1}{\Omega}\Sigma\left(|\nabla_{\mathrm{x}}S_h|^p + |\nabla_{\mathrm{y}}S_h|^p + |\nabla_{\mathrm{z}}S_h|^p + \epsilon\right)^{\frac{q}{p}}$, where $\Omega$ is the array size of the crystal function and $\nabla$ denotes the finite difference operations. $\epsilon$ is a small constant used to prevent singular gradient error. Here, we used $p = 2$ and $q = 1$ in our calculations, which is known as the Euclidean norm. The 2D coherent X-ray diffraction patterns for each minibatch were obtained by a random permutation at each iteration. This optimization seeks a solution that fits the dynamical diffraction model but also has a limited total variation for the reconstructed crystal function.

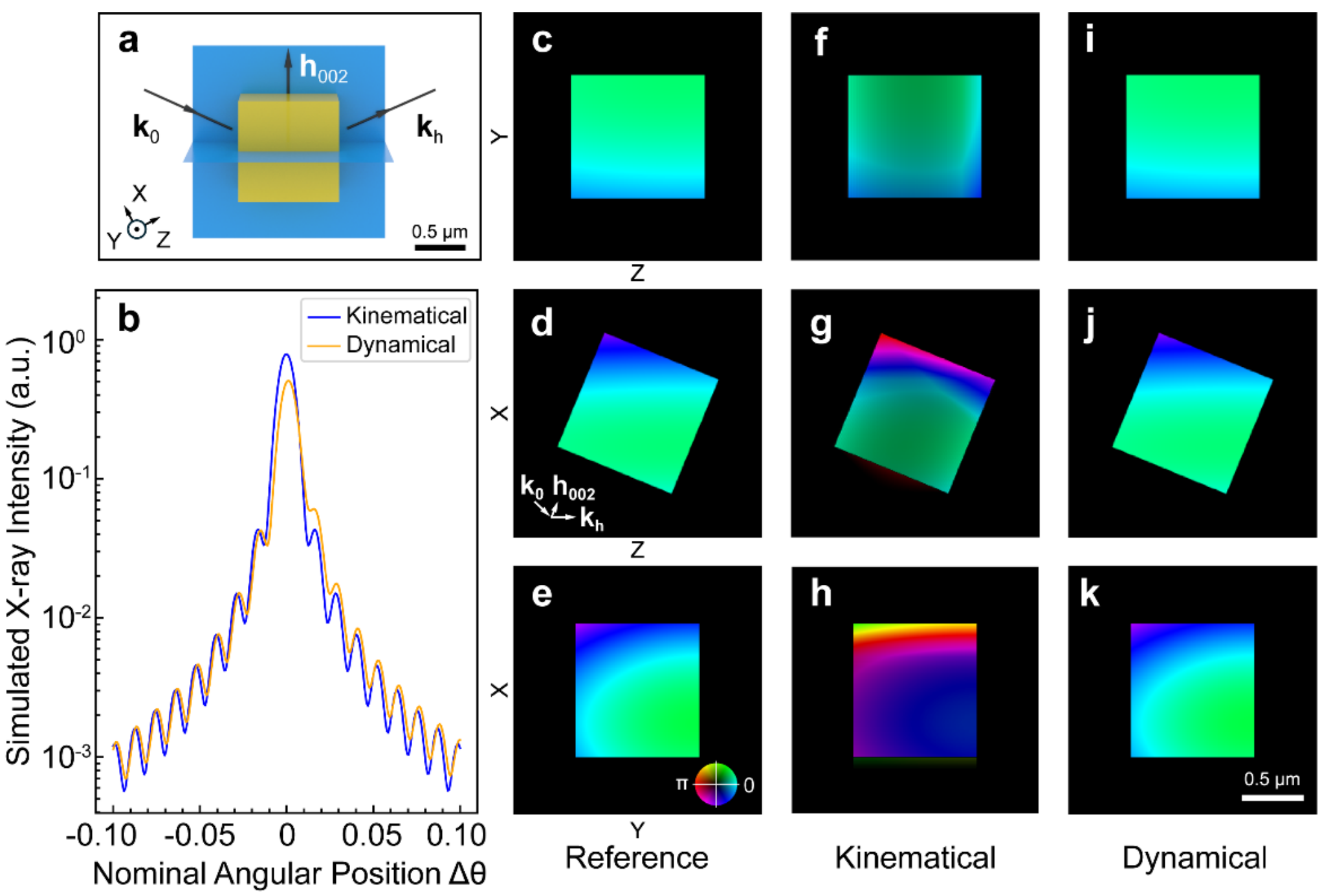

**Fig. 2 | Coherent X-ray dynamical diffraction effects.** (a) Schematic view of the diffraction geometry at sample. (b) Rocking curves as a function of crystal rotation angle, on a logarithmic scale, under the kinematical and dynamical approximation respectively. (c)-(e) Central X- Y- and Z-slices of the assumed complex crystal function. (f)-(h) Central X- Y- and Z-slices of the complex crystal function reconstructed with a traditional kinematical iterative method. (i)-(k) Central X- Y- and Z-slices of the complex crystal function reconstructed with the proposed dynamical method. The HSV color scheme, indicated in panel (e), uses brightness for the complex amplitude signal and a color wheel for the phase.

**Performance on simulated data**

The effects of coherent X-ray dynamical diffraction on these kinematical-diffraction-based inverse methods for reconstructing the deformation field can be illustrated with a numerical study. To highlight the feasibility of our proposed BCDI approach under coherent X-ray dynamical diffraction, a one-micron Au cube with lattice constant $a = 4.07$ Å was used to simulate the (002) BCDI rocking curve based on Eq. (3). The Au cubic unit cell is aligned along its cube edges. The cube is assumed to be deformed along its (002) crystal direction. Figure 2a shows a schematic of the diffraction geometry at sample with the incident X-ray $\mathbf{k}_h$ along the Z axis. Figure 2b presents the rocking curve (*i.e.*, the total intensity) of the $I_j$ as a function of rotation angle, along with the corresponding rocking curve scan from the kinematical approximation using the same Bragg diffraction geometry. For the simulation, Figures 2(c)-(e) demonstrate the corresponding deformation field used. The difference between the kinematical approximation and dynamical theory simulation is clearly observed in the position and intensity with a more complex structure of the rocking curve profile observed from the dynamical theory simulation. Compared with the kinematical result, it can be seen that there is a displacement of the whole profile, particularly around the Bragg peak position, in the positive direction of the $\Delta\theta$ axis for the dynamical theory simulations. This is well known in the dynamical theory[33] and is caused by refraction. This results in a phase ramp in the reconstructed result, which is a property of the Fourier transform.

To demonstrate how dynamical diffraction affects the reconstruction of the strained Au cube structure, the simulated Bragg Coherent X-ray diffraction patterns (based on dynamical theory) were inverted using the established error reduction (ER) and hybrid input–output (HIO) algorithms (see Methods for more details)[13,14,34,35]. Central cross-sectional views of the reconstructed crystal function are depicted in Figs. 2(f)-(h). Compared with the reference crystal function shown in Figs. 2(c)-(e), the reconstructed crystal function has a nonuniform amplitude profile, and the phase profile is considerably different, particularly at the top of the reconstructed particle, where the beams have travelled furthest. These differences are caused by the coherent X-ray dynamical diffraction effects. It has been proposed[20] that applying a linear approximation can remove these artifacts from refraction effects (see Fig. S1 for the corresponding corrected reconstruction). While this would be a satisfactory first approximation for weak phase effects, the coupling between the incident wave and the diffracted wave inside the crystal is generally strong and complicated, resulting in non-linear contributions. Small residual aberrations, apparently linked to the dynamical effects and not to refraction[15], can be attributed to the imaginary parts of the Fourier components of the susceptibility $\chi_h$ and $\chi_{\bar{h}}$, which introduce a small phase shift whenever the wave is reflected by a crystalline plane. This motivates use of a rigorous coherent X-ray dynamical diffraction model as an enhancement to traditional BCDI reconstruction. When this is applied to the model diffraction pattern in Figs. 2(i)-(k), a fairly constant crystal function profile, with distortions in excellent agreement between the original model and calculated results, is obtained. These results demonstrate the feasibility of using BCDI under coherent X-ray dynamical diffraction conditions. Our new computational method allows us to extend the accessible size of crystals to the microscale.

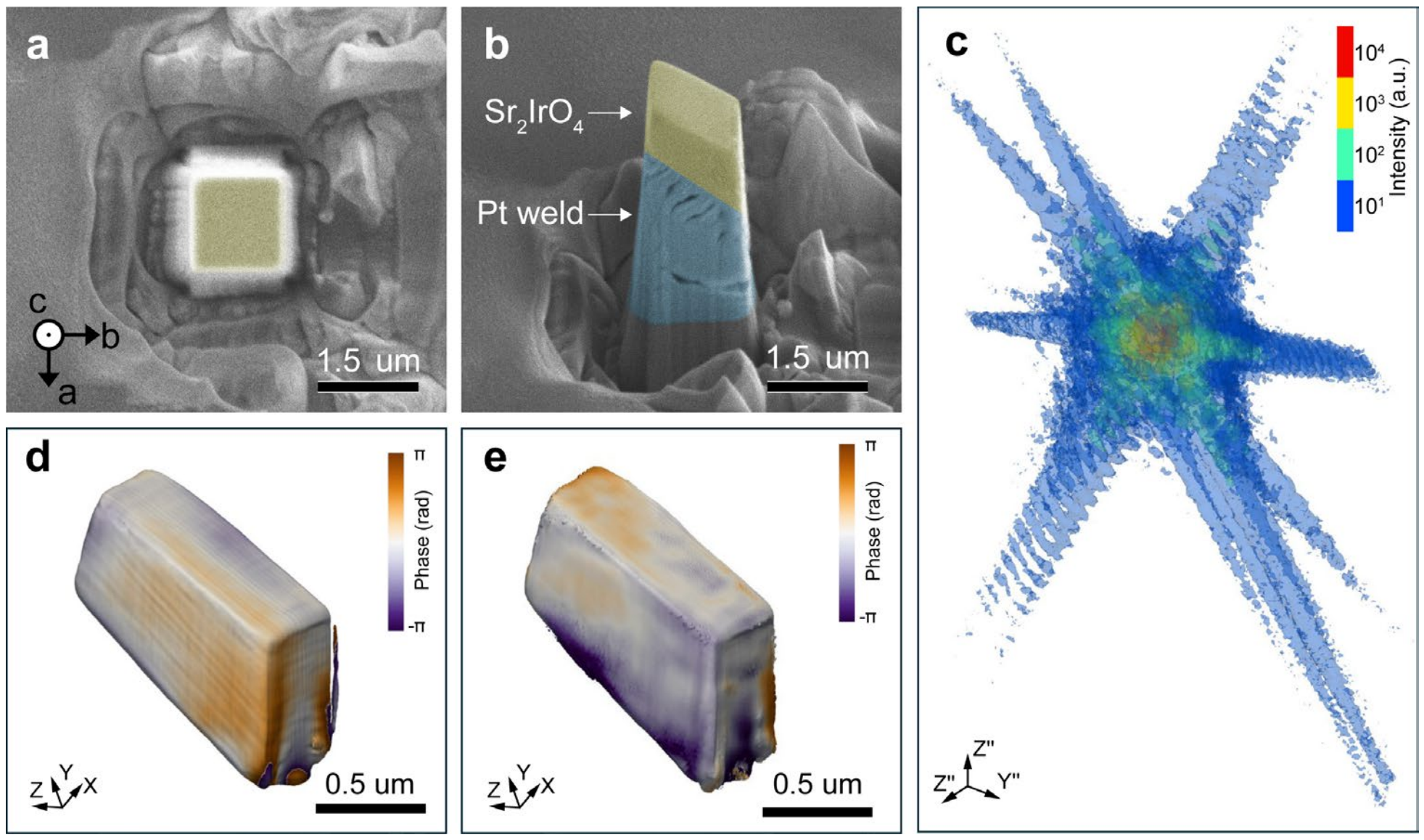

**Fig. 3 | BCDI experiment and sample characterization.** (a) Scanning electron microscopy image of the $Sr_2IrO_4$ sample viewed from the top, with the crystal orientation indicated by the black arrows. (b) Corresponding side view. (c) 3D isosurface plot of the measured coherent diffraction pattern in the detector frame from the (116) peak of the crystal. (d) 3D isosurface of the reconstructed crystal using the traditional kinematical iterative method. The crystal phase at the surface is illustrated by the color scale. (e) 3D isosurfaces of the reconstructed crystal using the proposed dynamical BCDI reconstruction method.

## Performance on experimental data

To quantitatively evaluate experimental BCDI data to see the benefits of modeling the coherent X-ray dynamical diffraction in the reconstruction process, a micron-sized $Sr_2IrO_4$ crystal was fabricated using the Focused Ion Beam (FIB) lift-out method. The lattice constants of the crystal are $a = b = 5.49$ Å and $c = 25.80$ Å with tetragonal symmetry. During the preparation, a large $Sr_2IrO_4$ crystal was preoriented crystallographically using a Laue diffractometer, and then a block with size of ~1.5×1.5×1.2 $\mu m^3$ was cut out using FIB milling. Afterwards, the obtained $Sr_2IrO_4$ crystal was welded with Pt onto a silicon wafer before final polishing, as shown in Figs. 3(a) and 3(b). The prepared crystal has a well-defined shape close to a parallelepiped with its crystal orientation indicated by the arrows in Fig. 3(a). This procedure follows our previous work[36]. Experimental BCDI measurements of the $Sr_2IrO_4$ sample were carried out at the 34-ID-C beamline at the Advanced Photon Source of Argonne National Laboratory at an incident X-ray energy of 10 keV. 3D coherent diffraction data were collected by rocking the sample. The crystal (116) peak was studied because of its relatively large structure factor. Figure 3(c) shows 3D isosurfaces of the Bragg coherent diffraction patterns of the (116) peak from the prepared $Sr_2IrO_4$ crystal, recorded in the detector coordinate X′′Y′′Z′′ at four isosurface levels. The pattern exhibits well-defined fringe spacings along streaks in directions perpendicular to the well-defined parallel facets of the crystal in real space. However, the 3D diffraction pattern is asymmetrical, suggesting the presence of strain modulation inside the crystal and/or dynamical effects.

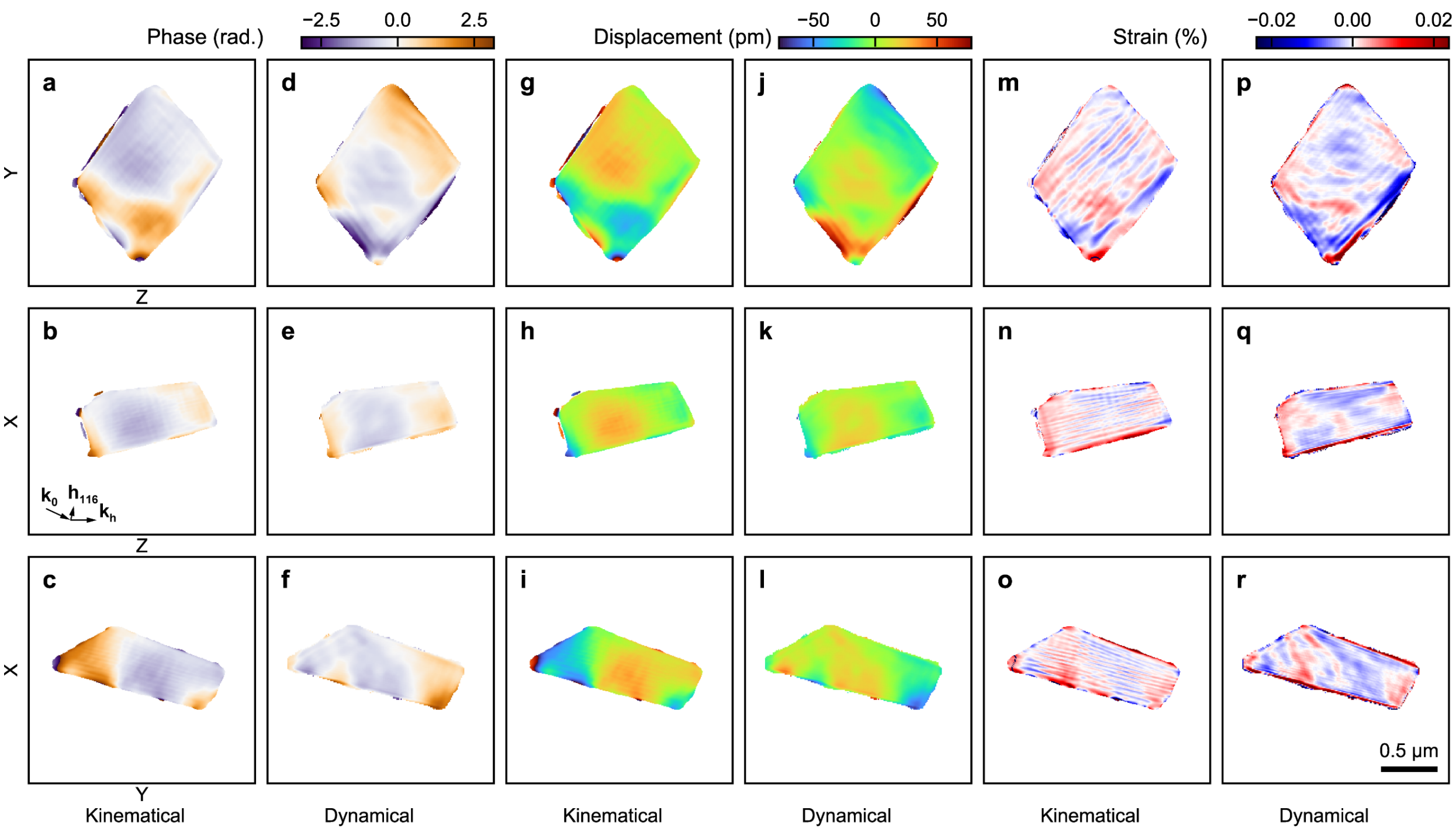

**Fig. 4. | Comparison between traditional kinematical iterative method and the proposed dynamical method.** (a)–(c) Central slices of retrieved phase using the traditional kinematical iterative reconstruction method. (d)-(f) Corresponding retrieved phase from the proposed method. (g)-(i) Central slices of obtained lattice displacement using the traditional kinematical method. (j)-(l) Corresponding results from the proposed method. (m)-(o) Central slices of obtained strain, the derivative of displacement with respect to position along the direction of the Q-vector, using the traditional method. (p)-(r) Corresponding results from the proposed method.

We inverted this 3D Bragg coherent diffraction pattern using the traditional iterative method, yielding results in Figs. 4(a)-(c) showing the extracted phase information from the reconstruction, where the phase is only allowed to span from $-\pi$ to $\pi$. It should be noted that the results have been converted into the lab coordinate system to remove the shear distortion of the detector geometry. For comparison, Figures 4(d)–(f) show the corresponding reconstructions with the proposed dynamical BCDI reconstruction approach, using the same 3D coherent diffraction data.

As shown in Fig. 4, both methods exhibit good agreement on the crystal shape, however, there is a significant difference between the phases. It can be clearly seen that the experimental image phases, reconstructed by the traditional kinematic approach, mix together the physical crystal strains with the artefacts of dynamical effects. For example, it can be seen that Figs. 4(h) and (k) appear to be almost identical, while Figs. 4(g) and (j) show a significant phase difference in the bottom corner of the crystal. This corner is the part of the crystal in contact with the Pt weld used to mount it, seen in Fig. 3(b), where contact strain is indeed expected. We can see that the displacement of the corner is much stronger in the dynamical image of Fig. 4(i) than the kinematical image of Fig. 4(l). The crystal displacement has been mostly cancelled out by the dynamical effects in Fig. 4(i) and is more accurately revealed in Fig. 4(l). Meanwhile, the rest of the crystal shows less phase variation in the dynamical-corrected image, suggesting that it is mostly dynamical artefact. The crystal corner also appears strongly in the strain image in Fig. 4(p). A similar result was seen previously in a BCDI image of a Pb crystal supported on a Si substrate[9], at that time, it was found that correcting for the X-ray refraction effects alone enhanced the apparent strain induced by the substrate but flattened the displacements in the rest of the crystal[20]. The dynamical theory indeed accounts for the refraction effects but should be more precise in general, for example accounting more completely for the X-ray optical propagation through strained regions of a crystal.

This discrepancy directly shows the difference between the kinematical and dynamical approaches. The coherent X-ray dynamical diffraction effect redistributes the corresponding reconstructed phase information; if this were not corrected, it could lead to mistakes in interpreting the obtained images, which could mislead subsequent analyses, such as displacement and strain analysis as presented in Figs. 4(g)-(r). Since the proposed method models the dynamical diffraction theory throughout the calculation, the dynamical artefacts are naturally excluded from the final image. Thus, the proposed method opens a path for accurate reconstruction of internal deformation fields in large crystals with nanoscale resolution.

## Discussion

The introduced BCDI reconstruction algorithm, based on the automatic differentiation (AD) framework of the Takagi-Taupin two-beam formalism, has been shown to be a powerful tool for the accurate reconstruction of deformation field in crystals undergoing coherent X-ray dynamical diffraction. This approach holds significant potential for probing large crystals, which have previously remained inaccessible due to the challenges associated with solving dynamical diffraction equations. Unlike traditional iterative methods, which directly use the 3D diffraction pattern obtained from a stack of measured 2D diffraction patterns for reconstruction, the current method employs a minibatch approach, utilizing subsets of data for each update. This strategy paves the way for future adaptations, such as incorporating rocking angle fluctuations and addressing uncertainties in optimization. Furthermore, the multipeak BCDI approach[37], which has great potential for obtaining full strain tensor information, would benefit from further refinements to the model. However, since the current framework relies on automatic differentiation, the inclusion of additional physical models may demand increased computational resources. Consequently, a careful balance between accuracy and efficiency must be considered to optimize the algorithm's performance. It should also be mentioned that the current TTEs used are only valid with pure $\sigma$ or $\pi$ polarization. We believe the proposed dynamical diffraction modeling approach will perform better by further considering the effects of arbitrarily polarized incident X-rays[38]. However, without restricting its generality, it can be overcome experimentally by tuning the polarization direction of the incident X-ray or sample geometry during a BCDI experiment. For unpolarized incident X-ray, we solve the TTEs for σ component and π component, separately. The final coherent x-ray diffraction intensity is found by summing the intensities from these two orthogonal polarization states.

Despite these challenges, the proposed method represents a significant step forward in the field, offering new opportunities for studying complex deformation fields in large crystals with unprecedented precision. As the fourth-generation synchrotron radiation facilities bring large gains in X-ray source coherence, the approach presented is expected to find important applications for looking at strain in micron-scale crystals.

In summary, we have developed a dynamical diffraction modeling approach that enables precise visualization of internal deformation fields measured by coherent X-ray dynamical diffraction, significantly advancing the characterization of large single crystals. By integrating Bragg coherent diffraction imaging (BCDI) with dynamical diffraction modeling, our method is not only compatible with modern optimization techniques but also extends the accessible size range of crystals. A key innovation lies in the direct incorporation of the TTEs on an orthogonal grid into the BCDI reconstruction framework, which effectively eliminates the distortions of the dynamical diffraction in the resulting images. This approach allows for the reconstruction of deformation fields in large crystals with nanoscale resolution, unlocking unprecedented opportunities to explore previously inaccessible structural phenomena in bulk crystalline materials. The inclusion of an ATV constraint further enhances the robustness and applicability of our approach, particularly in handling complex strain fields. Our method demonstrates exceptional potential for fully exploiting the capabilities of enhanced coherence and flux expected from fourth-generation synchrotron X-ray facilities. We anticipate that this work will catalyze the widespread adoption of BCDI at next-generation synchrotrons, enabling transformative studies in materials science. This advancement not only addresses current limitations in crystal deformation field characterization but also establishes a foundation for future investigations of strain-mediated phenomena in functional materials.

## Methods

### BCDI Simulation

In our simulations, we considered the incident plane wave with 8 keV photon energy and (002) reflection conditions of a cube-shaped block of Au. For the (002) reflection, a $\pi$ polarized diffraction geometry is assumed, given as $\chi_0 = -0.94070 \times 10^{-4} + i0.84962 \cdot 10^{-4}$ and $\chi_h = 0.52548 \times 10^{-4} - i0.59889 \cdot 10^{-5}$. The extinction lengths are 354 nm and 2710 nm for the Bragg and Laue geometries, respectively[18,39,40]. The corresponding *Pendellösung* distance[33] is given as 863 nm. In this diffraction geometry, the Bragg angle in this condition is $2\theta_{\mathrm{B}} = 44.76°$ ,where $2\theta_{\mathrm{B}}$ is the Bragg angle formed by $\mathbf{k}_0$ and $\mathbf{k}_h$. The rocking scan was simulated covering the angular range from $-0.2°$ to $+0.2°$, with an angular increment of $1.56 \times 10^{-3}$ degree. The scattered intensity $I_j$ in the far field was calculated with 256 sampling points in each direction.

### BCDI Reconstruction

When applying the phase retrieval algorithm, the Bragg 3D diffraction patterns were used as input to the iterative phase retrieval scheme to reconstruct the corresponding real-space structural information. During reconstruction, the initial support size of the particle in real space was half the size of the input diffraction pattern array in each dimension. The algorithm switched between hybrid input-output (HIO) with $\beta$ = 0.9 and error reduction (ER) after every 50 iterations. After the first 100 iterations, the shrink-wrap method[41] was applied in real space to dynamically update the support every ten iterations. At the end of the reconstruction, 200 steps of error reduction were applied. The total number of iterations was 1800. After reconstruction, all results were converted from the detector to sample coordinates. The computation was performed on a system with 251 GB of RAM and 8 NVIDIA-SMI A100 GPUs.

The proposed algorithm for BCDI under coherent X-ray dynamical diffraction was implemented using PyTorch[28], where the Wirtinger calculus was applied for gradient calculation of the complex-valued array. The same computational resource has been used for the proposed algorithm. To speed up the reconstruction process, we used the reconstructed result from the iterative method in the Cartesian Laboratory coordinates to initialize the amplitude function and deformation field for the proposed algorithm. For the (116) reflection of $Sr_2IrO_4$, a $\pi$ polarized diffraction geometry was approximated to carry out the calculation, given as $\chi_0 = -0.24947 \times 10^{-4} + i0.96953 \times 10^{-4}$ and $\chi_h = 0.13373 \times 10^{-4} - i0.84241 \times 10^{-6}$. The extinction lengths are 633 nm and 9036 nm for the Bragg and Laue geometries, respectively. The corresponding *Pendellösung* distance is given as 2876 nm. To minimize the difference between the experimental diffraction patterns, the intensity scale factor applied for the calculated diffraction pattern was also optimized. During the reconstruction, the BCDI reconstructions were completed using the Adam optimizer[42]. The learning rate was initialized to 0.15 and was dynamically reduced by the scheduler using the loss metrics quantity when no improvement is seen for a "patience" number of epochs. For each epoch, the experimental coherent diffraction was divided randomly into five groups. Inspired by the traditional iterative method, the shrink-wrap method is also applied during the optimization after every 10 epochs. The reconstruction uses 6 hours to run 600 iterations.

**Sample Preparation**

The high-quality $Sr_2IrO_4$ single crystal was grown from off-stoichiometric quantities of $SrCl_2$, $SrCO_3$, and $IrO_2$ using self-flux techniques, described in a previous publication[36,43]. A large $Sr_2IrO_4$ crystal was pre-oriented crystallographically using a Laue diffractometer. The Focused Ion Beam (FIB) lift-out protocol was employed to extract a micron-sized $Sr_2IrO_4$ sample from the aligned large crystal. Using a Helios G5 Dual Beam SEM/FIB Microscope, a ~5×5×5 $\mu m^3$ sample was extracted from the bulk $Sr_2IrO_4$ using an Omniprobe needle, and the cube-shaped sample was attached to a silicon wafer with a platinum weld. The lifted $Sr_2IrO_4$ sample was then trimmed down to approximately ~1.5×1.5×1.2 $\mu m^3$ in each direction by gradually reducing the ion beam voltages and currents. Finally, a cleaning step with a 1 keV Ga beam was used to remove surface damage from the previous FIB milling steps.

**BCDI Experiments**

The Bragg coherent X-ray diffraction experiments were carried out at the 34-ID-C beamline of the Advanced Photon Source. A double crystal monochromator was used to select the energy of 10.0 keV and Kirkpatrick–Baez mirrors were used to focus the X-ray beam. To fully illuminate the crystal for valid Bragg coherent X-ray imaging, the beam size before focusing was adjusted to 1.2 × 1.4 $\mu m^2$ (H×V) by changing the coherence-defining entrance X-ray slit to 10 × 20 $\mu m^2$ (H×V). The sample was mounted on a stage for BCDI measurements. Since the $Sr_2IrO_4$ sample was pre-aligned before FIB preparation, the precise crystal alignment was quickly determined by using the (204) and (116) peaks of the sample, with a six-axis diffractometer to maneuver the sample orientation. After the $Sr_2IrO_4$ crystal was aligned, the BCDI data were collected by rocking the sample around the (116) peak. The corresponding diffraction signal was collected by a Timepix photon-counting detector mounted 2.5 m away from the $Sr_2IrO_4$ sample, and a vacuum flight tube was used to avoid air scattering contribution to the diffraction signal.

**Lattice Strain Calculation:**

For a given Bragg peak, $\mathbf{h}$, there is a simple linear relationship between the observed phase $\varphi_h(\mathbf{r})$ and the crystal displacement field $\mathbf{u}(\mathbf{r})$: $\varphi_h(\mathbf{r}) = -\mathbf{h} \cdot \mathbf{u}(\mathbf{r})$. The strain is related to the variation of the d-spacing of the $Sr_2IrO_4$ crystal based on the measured Bragg peak. For the $Sr_2IrO_4$ crystal (116) peaks, the strain can be calculated as: $\epsilon_{116} = \partial u_{116} / \partial d_{116}$, where $u_{116}$ is the corresponding displacement field and $d_{116}$ is the real-space coordinate perpendicular to the (116) plane.

**Acknowledgements**

We thank Ivan Vartanyants, Anatoly Shabalin, Hanfei Yan and Yuan Gao for helpful discussions of dynamical diffraction formalism. Work at Shanghai Advanced Research Institute was funded by the '100 Talents Project' of the Chinese Academy of Sciences. Work at Brookhaven National Laboratory was supported by the U.S. Department of Energy, Office of Science, Office of Basic Energy Sciences, under Contract No. DE-SC0012704. Measurements were conducted at the Advanced Photon Source (APS) beamline 34-ID-C, which was supported by the U. S. Department of Energy, Office of Science, Office of Basic Energy Sciences, under Contract No. DE-AC02-06CH11357. The beamline 34-ID-C was built with U.S. National Science Foundation grant DMR-9724294.

**Author Contributions**

L.W. developed the dynamical model and performed the data analysis with the support from I.K.R., S.Y. and A.L.. L.W., I.K.R., D.Y., R.J.H. and W.C. performed experimental BCDI measurements. W.W. prepared the FIB sample. L.W. and I.K.R. wrote the manuscript and all the authors contributed to the discussion of the manuscript.

**Data availability**

The data that supports the findings of this study are available from the corresponding authors upon request.

**Code availability**

All codes that support the findings of this study will be available upon request to the corresponding authors.